\documentclass[11pt,a4paper]{article} 

\usepackage{anysize}
\usepackage[format = hang]{caption}
\usepackage{amsmath,amsthm,verbatim,amssymb,amsfonts,amscd, graphicx}
\usepackage{graphics,natbib}
\usepackage{titlesec,footmisc}
\usepackage{hyperref} 

\usepackage{listings}
\usepackage{booktabs} 

\theoremstyle{plain}

\theoremstyle{definition}

\newcommand{\corr}{\text{corr}}

\def\E{{\rm E}}
\def\Var{{\rm Var}}

\def\tr{{\rm t}}
\def\midd{\,|\,}

\def\dd{{\rm d}}
\def\hatt{\widehat}
\def\arr{\rightarrow}
\def\N{{\rm N}}
\def\half{\hbox{$1\over 2$}}

\def\binom{{\rm Bin}}

\def\se{{\rm se}} 

\def\beq{\begin{eqnarray}}
\def\eeq{\end{eqnarray}}

\def\beqn{\begin{eqnarray*}}  
\def\eeqn{\end{eqnarray*}}

\def\E{{\rm E}}
\def\Var{{\rm Var}}

\def\dd{{\rm d}}
\def\N{{\rm N}}

\def\Pr{P}

\def\quadandquad{\quad {\rm and} \quad}
\def\arr{\rightarrow}
\def\hatt{\widehat}

\def\sumin{\sum_{i=1}^n}

\def\half{\hbox{$1\over2$}}

\def\rootn{\sqrt{n}}

\def\midd{\,|\,}
\def\tr{{\rm t}}
\def\dell{\partial}

\def\prof{{\rm prof}}

\def\true{{\rm true}}

\def\pois{{\rm Pois}}

\def\cc{{\rm cc}}

\titleformat{\section}{\normalfont\large\sc\centering}{\thesection}{1em}{}
\titleformat{\subsection}[runin]{\normalfont\large\bfseries}{\thesubsection}{1em}{}
\numberwithin{equation}{section} 
\renewenvironment{abstract}
               {\list{}{\rightmargin\leftmargin}%
                \item[\text{\hspace{10mm}\sc Abstract.}]\relax}
               {\endlist}

\begin{document}

\def\heute{May 17, 2026}

\begingroup
\begin{centering} 
  \Large{\bf Modelling pairs of Poissons and binomials \\
    with negative correlation}\\[0.8em]  
\large{\bf Nils Lid Hjort} \\[0.3em] 
\small {\sc Department of Mathematics, University of Oslo} \\[0.3em]
\small {\sc {\heute}}\par
\end{centering}
\endgroup


\begin{abstract}
\small{Suppose $f_1(x)$ and $f_2(y)$ are given marginals
  for pairs $(x,y)$. I consider the construction 
  $f_1(x)f_2(y)\{ 1+\alpha h_1(x)h_2(y) \}$,
  where $h_1$ and $h_2$ are seen as bounded adjustment functions,
  normalised to have means zero under $f_1$ and $f_2$. 
This defines a bivariate distribution for $(X,Y)$
with the specified marginal densities $f_1$ and $f_2$,
with an interval of permissible values of $\alpha$,
both positive and negative; in particular, independence
corresponds to an innter point in the adjustments parameter
region. Applications to bivariate Poisson distributions,
allowing both positive and negative correlation, 
are discussed. As illustration I provide a more accurate
and extended analysis of a Poisson pairs dataset,
pertaining to competing seeds and plants, for $n=958$
plots of soil, earlier analysed in the well-cited paper 
Lakshminarayana, Pandit, Rao, Srinivasa (1999).
The general apparatus is also shown
to work for negatively correlated binomials.
Those methods are illustrated in a meta-analysis
framework for two-by-two tables across different studies,
pertaining to the Audit-C screening questionnaire
for alcohol use disorders, where again negative correlation
is demonstrated, between $X$, the number of correct `yes', 
and $Y$, the number of correct `no'. 


}
\end{abstract}



\section{A general bivariate construction}
\label{section:intro}

Suppose $f_1(x)$ are $f_2(y)$ are given marginal probability
densities, and let $g_1(x)$ and $g_2(y)$ be bounded
functions; these will be called adjustment functions below. 
For notational convenience I first let these
be ordinary densities, i.e.~with respect to Lebesgue measure,
but matters generalise easily to the discrete setting,
as we shall see later with the Poisson and binomial distributions.

\subsection{The one-parameter deviation from independence.}

Consider 
\beq
f(x,y)=f_1(x)f_2(y)\{ 1+\alpha h_1(x)h_2(y)\},
\label{eq:1.1}
\eeq 
with $h_1(x)=g_1(x)-a_1$ and $h_2(y)=g_2(y)-a_2$ the
adjustment functions normalised to have means zero
under $f_1$ and $f_2$, i.e. 
\beqn
a_1=\E\,g_1(X)=\int f_1(x)g_1(x)\,\dd x 
   \quadandquad 
   a_2=\E\,g_2(Y)=\int f_2(y)g_2(y)\,\dd y.
\eeqn 
This defines a bivariate density for $(X,Y)$, 
with marginals $f_1(x)$ and $f_2(y)$, provided 
$\alpha$ is small enough to secure $\alpha h_1(x)h_2(y)$
being inside $(-1,1)$ for all $(x,y)$. Let in fact 
\beqn
c_1=\max_x |g_1(x)-a_1| \quadandquad
c_2=\max_y |g_2(y)-a_2|,
\eeqn 
both finite by assumption. 
Then $|\alpha h_1(x)h_2(y)|\le|\alpha|c_1c_2$ everywhere, so as long as 
\beq
|\alpha|< 1/(c_1c_2),
\label{eq:1.2}
\eeq 
$f(x,y)$ of (\ref{eq:1.1}) is a bona fide density. We note that
both positive and negative $\alpha$ are allowed,
and in particular that $\alpha=0$, corresponding to independence,
is an inner point.  

Let $X$ and $Y$ have means $\mu_1$ and $\mu_2$
and standard deviations $\sigma_1$ and $\sigma_2$. 
We then have 
\beqn
\E\,XY=\mu_1\mu_2+\alpha\nu_1\nu_2,
\eeqn 
where $\nu_1=\int xf_1(x)h_1(x)\,\dd x$ and
$\nu_2=\int yf_2(y)h_2(y)\,\dd y$. 
which leads to the correlation formula 
\beq
\corr(X,Y)=\alpha{\nu_1\nu_2\over \sigma_1\sigma_2}.
\label{eq:1.3}
\eeq 
  
It is not entirely clear from the outset which constructions 
can be judged as more fruitful than others. We would like 
the $f(x,y)$ of (\ref{eq:1.1}) to include a decent amount
of extra changes, compared to independence, which
means that $g_1$ and $g_2$ should exhibit some
range of expression -- but not too much, since that 
would limit the range of admissible $\alpha$,
by (\ref{eq:1.2}). Also note that scaling constants might be
moved between $\alpha$ and the $g_1$ and $g_2$ functions
in (\ref{eq:1.1}); if we multiply $g_1$ and $g_2$ 
by a constant $k$, this produces the same model,
with the previous $\alpha$ replaced by $\alpha/k^2$.

The conditional distributions are easy to read off
from the basic construction. A given $X=x$ influences $Y$,
and similarly for $Y=y$ influencing $X$, via 
\beq
\begin{array}{rcl} 
f_1(x\midd y)
   &=&\displaystyle f_1(x)\,\{1+\alpha h_1(x)h_2(y)\}, \\ 
f_2(y\midd x)
   &=&\displaystyle f_2(y)\,\{1+\alpha h_1(x)h_2(y)\}, 
\end{array} 
\label{eq:conditional}
\eeq 
demonstrating the adjustments to their marginals
in view of extra information. 

\subsection{Modelling, interpretation, inference.}

First one may note that the (\ref{eq:1.1}) is of the copula type,
related, in its simplest form, to the joint density
$f_0(u,v)=1+\alpha h(u)h(v)$ on the unit square, with the $h(u)$
function constructed to have zero mean under uniformity;
the range of $\alpha$ is determined by that of the $h(u)$. 
As such the basic construction worked with in this
article is inside classical copula terrain;
see e.g.~Nelson (1999) for a general introduction,
and consult Mikosch (2006) for a lively discussion. 
The novel aspects introduced in the present contribution are
(i) its use for negative correlated Poisson pairs
and binomials, i.e.~with discrete distributions, where
generic copula models have difficulties;
(ii) to allow for extra fine-tuning parameters inside
the adjustment functions, for potentially better fit to data;
(iii) building alternative bivariate dependence models with
normal marginals; and 
(iv) the construction of models for bivariate Poisson regression
and bivariate binomial regression. 

In a statistical scenario, with data pairs
$(x_1,y_1),\ldots,(x_n,y_n)$ with known or perhaps well
estimated marginals, it would be easy to estimate
the dependence parameter via likelihood methods, maximising
\beqn
\ell_n(\alpha)=\sumin \log\{1+\alpha h_1(x_i)h_2(y_i)\}
\eeqn
to get $\hatt\alpha$. Confidence intervals and curves
can also be put up, via the standard result 
$\rootn(\hatt\alpha-\alpha)\arr_d\N(0,J(\alpha)^{-1})$,
with the inverse Fisher information matrix; also, one
sees that $J(0)=1$, so an easy test for independence
is $|\rootn\hatt\alpha|>z_0$, the normal quantile for
the appropriate significance level. 
This test is essentially the most powerful one for
detecting deviations from independence in the direction
of $f(x,y)/\{f_1(x)f_2(y)\}=1+\alpha h_1(x)h_2(y)$.
This suggests that good choices for the adjustment functions
should reflect notions of which $h_1(x)h_2(y)$ can be
anticipated to be present. 

Estimation carried out as pointed to here, via
`marginals first, then correlation', can be called
two-stage estimation, whereas the more principled
and typically better approach is to apply likelihood
methods to the full model;
see Ko and Hjort (2019), Ko, Hjort, and Hob\ae k Haff (2019),
for further discussion and pertinent examination regarding this. 


For statistical modelling we might work with classes of such
adjustment functions $g_1$ and $g_2$ and 
let the data influence how to fine-tune these. 
An example is provided below, for a bivariate Poisson
model, where $g(x)=\exp(-tx)$ is used, and where 
data inform us on which fine-tuning values $t$
do better than others, in terms of fitting the data well. 



Details for building such correlated Poissons (and, in particular,
allowing negative correlation) are in Sections \ref{section:poissonA},
with the application to the Indian seeds-and-plants dataset
in Section~\ref{section:poissonB}; seeds and plants compete
for the same soil. 
Similarly, details for correlated binomials are worked
through in Sections \ref{section:binomialsA}, with
the application to the Audit-C test questionnaires
in Section \ref{section:binomialsB}. Here both $X$,
the number of `yes' out of the yes individuals,
and $Y$, the number of `no' out of no, are binomials,
with negative correlation, as it turns out. 
It is also instructive
to see such bivariate models built around normal
marginals, where we learn in Section \ref{section:normals}
that several constructions lead to bivariate models,
with uncorrelated normal marginals.
The article is rounded off with a little list of
concluding remarks in Section \ref{section:concluding}.
I point to extensions to regression models,
say with negatively correlated Poisson or binomial
pairs, with correlations potentially depending
on covariates, and to other bivariate constructions,
e.g.~for exponentially distributed pairs. 

\section{A bivariate Poisson distribution}
\label{section:poissonA}

Consider the usual Poisson density function $f(x,\theta)$.
For $X$ having this distribution, consider 
the normalised adjustment function
\beq
h(x)=g(x)-a(\theta)=\exp(-t x)-\E\,\exp(- tX), 
\label{eq:gchoicepoisson}
\eeq 
for which 
\beqn
a(\theta)=\E\,g(X)=\sum_x f(x,\theta)g(x)=\exp(-d\theta), 
\quad {\rm with\ }d=1-\exp(-t). 
\eeqn 
Assuming the apparatus to be applied in situations
where the potential deviances from Poisson
are not very different,  we employ the same $h$ function for both,
and are led to the bivariate distribution
\beq
f(x,y)=f(x,\theta_1)f(y,\theta_2)
[1+\alpha\{g(x)-a(\theta_1)\}\{g(y)-a(\theta_2)\}
\label{eq:3.1}
\eeq
for $x,y=0,1,2,\ldots$. The marginals are Poisson
with parameters $\theta_1$ and $\theta_2$. Let 
\beqn
c(\theta)=\max_x |\exp(-tx)-a(\theta)|
   =\max\{1-a(\theta),a(\theta)\}
   =\max \{\exp(-d\theta),1-\exp(-d\theta)\}. 
\eeqn 
From (\ref{eq:1.3}), the range of allowed $\alpha$ is 
\beq
|\alpha|<{1\over c(\theta_1)c(\theta_2)},
\label{eq:3.2}
\eeq
with independence corresponding to the inner parameter point
$\alpha=0$. 

To compute the correlation, via (\ref{eq:1.3}), we need 
\beqn
\nu(\theta)
&=&\sum xf(x)\{g(x)-a(\theta)\} \\
&=&\theta\exp(-t)\exp[-\theta\{1-\exp(-t)\}]-\theta a(\theta)
   =-\theta a(\theta)d. 
\eeqn 
The correlation, with the $g(x)$ choice
of (\ref{eq:gchoicepoisson}), is hence
\beqn
\corr(X,Y)=\alpha \theta_1^{1/2}\theta_2^{1/2}
   a(\theta_1)a(\theta_2)d^2. 
\eeqn 
The bivariate Poisson model (\ref{eq:3.1}) hence
has correlation range 
\beqn
|\corr(X,Y)|\le r= {\theta_1^{1/2}\theta_2^{1/2}
  a(\theta_1)a(\theta_2)d^2 \over
  \max\{a(\theta_1),1-a(\theta_1)\}
       \max\{a(\theta_2),1-a(\theta_2)\}}. 
\eeqn 
For the case of equal Poisson rates,
\beqn
r=\theta d^2 \Bigl[{ a(\theta)\over \max\{a(\theta),1-a(\theta)\}}\Bigr]^2,
\quad {\rm with\ }a(\theta)=\exp(-d\theta). 
\eeqn 



\section{Seeds and plants: bivariate Poisson with negative correlation} 
\label{section:poissonB}

\begin{figure}[h]
\centering
\includegraphics[scale=0.36]{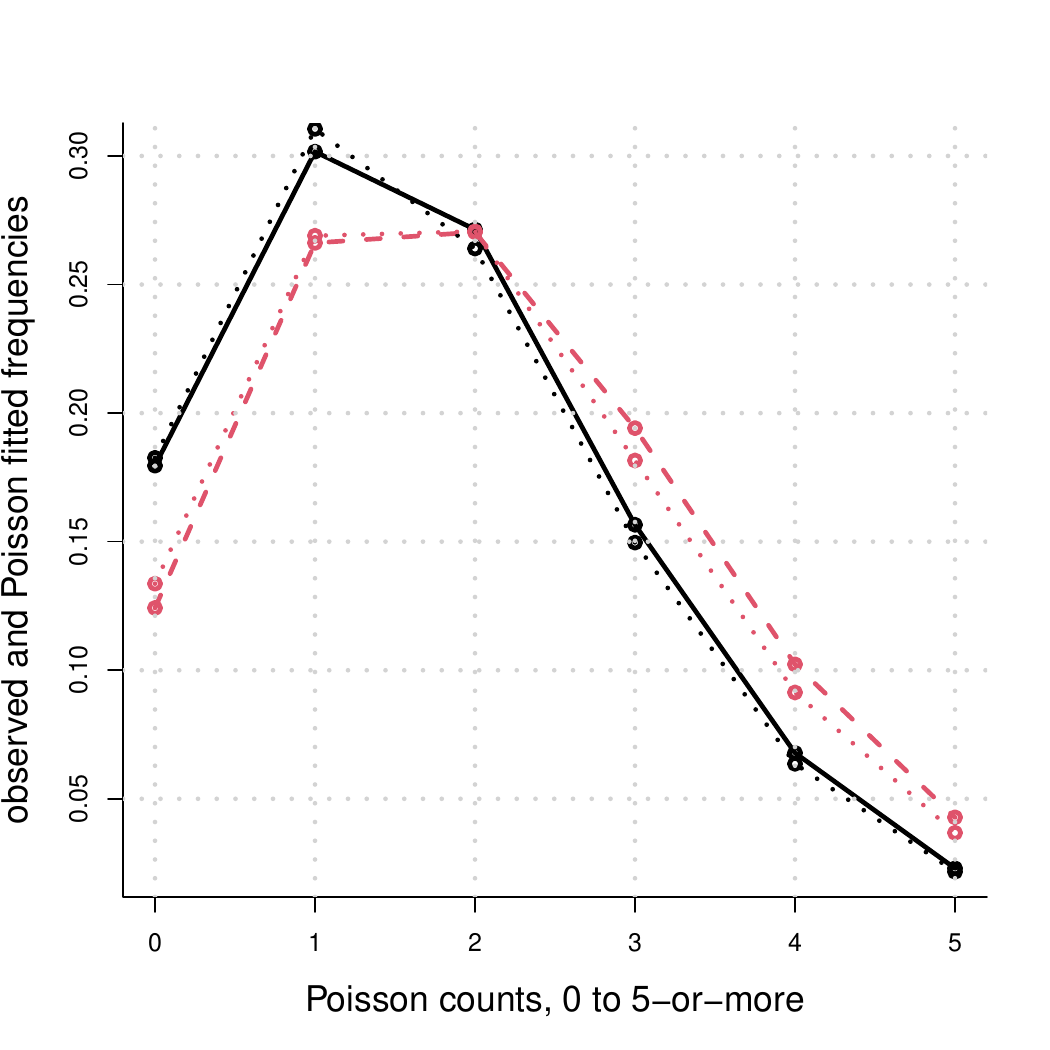}
\includegraphics[scale=0.36]{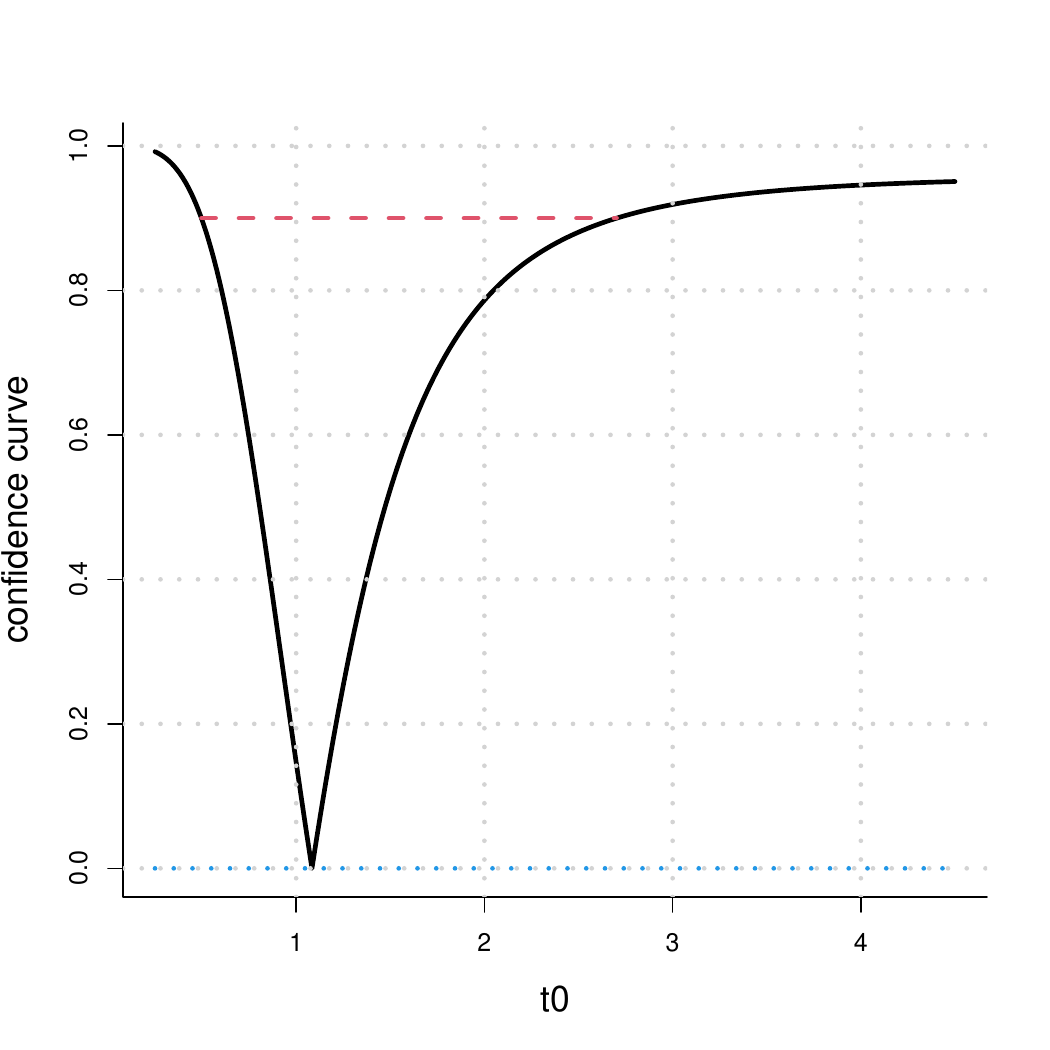}
\caption{For the seeds and plats dataset, with $n=958$
  negatively correlated Poisson pairs: 
  Left panel:
  Observed frequencies, for counts 0, 1, 2, 3, 4, 5-and-more,
  in full black and slanted red, 
  along with their associated Poisson estimates,
  as dotted curves, for seeds (top curves) and plants (lower curves).
  The Poisson estimated rates are
  $(\hatt\theta_1,\hatt\theta_2)=(1.700,2.0.12)$.
  Right panel: Confidence curve for the tuning parameter $t$,
  in the adjustment function $g(x)=\exp(-tx)$, 
  for the seeds-and-plants dataset. The point estimate is 1.084; 
  the 90\% confidence interval is $[0.497,2.272]$;
  and it levels off at level 0.96, i.e.~giving $t=\infty$
  a positive chance.}
\label{figure:fig14}
\end{figure}

Lakshminarayana, Pandit, Rao, Srinivasa (1999) considered
data from an experiment involving the number of 
seeds $x$ and the number of plants $y$ grown
over the number $n=958$ plots of size five square feet. 
The 958 pairs are conveniently summarised in a table of values
$(x,y)$, with $x$ and $y$ ranging trough $0,1,2,3,4,5$;
here `5' means `5-or-more';
see the top half of Table \ref{table:thedata}.
Computing frequencies $\hatt f_1(x)$ and $\hatt f_2(y)$
shows good agreement with the Poisson, see
Figure \ref{figure:fig14}, left panel, with separately
estimated Poisson rates
$(\hatt\theta_1,\hatt\theta_2)=(1.700,2.0.12)$.
Below we demonstrate that there is a negative correlation
between seeds and plants. 

\begin{table}
\begin{small}
\begin{verbatim}
         the data:                    
         0    1    2    3    4    5   
    0    7   41   54   40   21    9 
    1   36   79   73   58   30   13  
    2   39   70   69   47   25   10   
    3   24   41   39   26   14    6  
    4   10   18   18   11    6    2   
    5    3    6    6    4    2    1
        expected, independence:         expected, three-parameter model: 
         0    1    2    3    4    5      0    1    2    3    4    5
    0   26.1 52.5 52.8 35.4 17.8 10.5   16.1 50.1 56.9 39.8 20.3 12.0
    1   41.5 83.5 84.1 56.4 28.4 16.7   41.5 83.5 84.1 56.4 28.4 16.7
    2   33.0 66.4 66.9 44.8 22.6 13.3   37.6 67.6 65.0 42.9 21.4 12.6
    3   17.5 35.2 35.4 23.8 12.0  7.0   20.8 36.0 34.1 22.3 11.1  6.5
    4    7.0 14.0 14.1  9.5  4.8  2.8    8.4 14.4 13.5  8.8  4.4  2.6
    5    3.0  6.0  6.0  4.0  2.0  1.2    3.6  6.1  5.7  3.8  1.9  1.1
\end{verbatim} 
\end{small}
\caption{{\it Upper half:}
  The Indian seed- and plant-data,  for a total of $n=958$ pairs,
  sorted here with the number of pairs $(x_i,y_i)$
  for the given boxes of $x,y=0,1,2,3,4,5$,
  and where `5' means `5-or-more'.
  {\it Lower half:}
  The expected numbers, computed under
  respectively
  (i) the independent Poissons model (left),
  with  Poisson rates $1.700$ and $2.013$ for seeds and plants; 
  and (ii) the three-parameter correlation model (right). 
  The three-parameter model yields expected numbers
  much closer to the actual data.}
\label{table:thedata}
\end{table} 

For this application of our methodology we first 
work with the adjustment function $g(x)=\exp(-x)$,
i.e.~using $t=1$ when employing model (\ref{eq:3.1}).
That this is a sensible value for the present purposes
is argued for below, and related to the confidence curve
$\cc(t)$ for that tuning parameter shown in Figure \ref{figure:fig14},
right panel. The log-likelihood function 
\beqn
\ell(\theta_1,\theta_2,\alpha)
   =\sumin \bigl(\log f(x_i,\theta_1)+\log f(y_i,\theta_2) 
   +\log[1+\alpha\{g(x_i)-a(\theta_1)\}\{g(y_i)-a(\theta_2)\}]\bigr)
\eeqn
can be maximised, leading to the following 
maximum likelihood (ML) estimates, along with estimated standard deviations
(standard errors). These are somewhat different 
from estimates obtained in Lakshminarayana et al.~(1999), 
since they used what essentially amounts to moment estimators.
Also, they did not have methodology for assessing precision
of their estimators, whereas behaviour of our estimators 
can be accurately characterised via standard ML theory.
With $\gamma=(\theta_1,\theta_2,\alpha)$, 
and ML estimators $\hatt\gamma=(\hatt\theta_1,\hatt\theta_2,\hatt\alpha)$, 
we have $\hatt\gamma\approx_d\N_3(\gamma,\hatt J^{-1})$, 
in terms of the observed information matrix 
$\hatt J=-\dell^2\ell(\hatt\theta)/\dell\theta\,\dell\theta^\tr$.

\begin{figure}[h]
\centering
\includegraphics[scale=0.36]{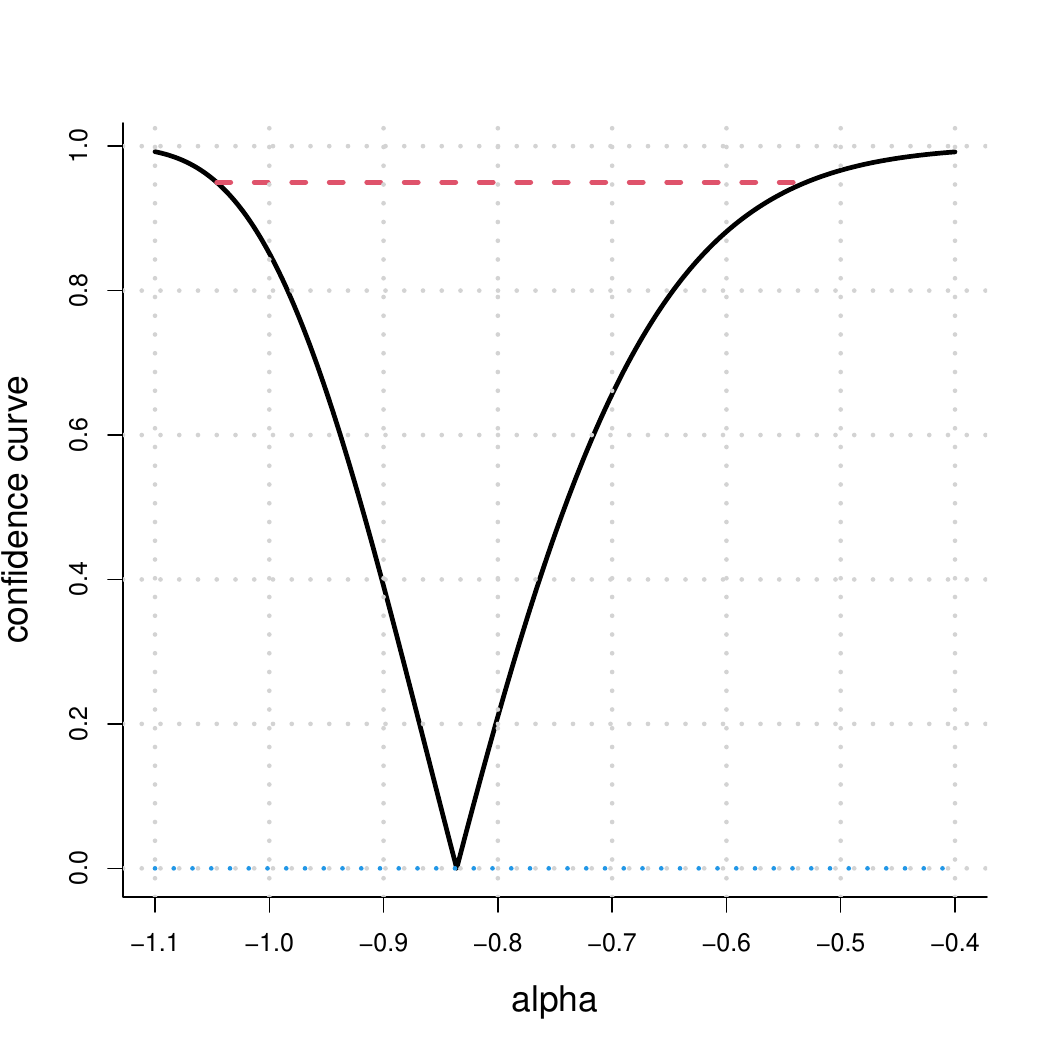}
\includegraphics[scale=0.36]{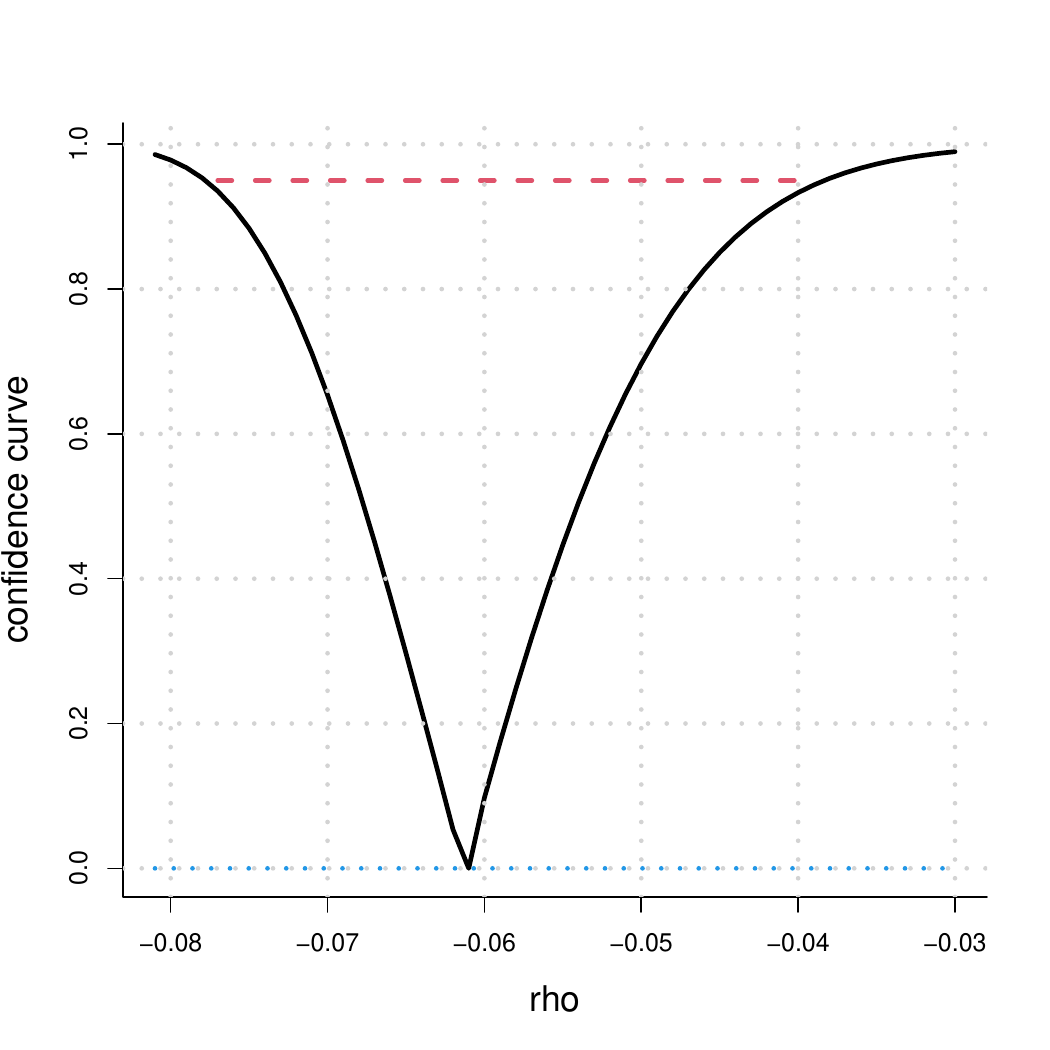}
\caption{Confidence curve for $\alpha$,
  then for the correlation $\rho$,  
  for the seeds-and-plants dataset.
  They are significantly negative, with 95 percent
  intervals $[-1.046,-0.531]$ and $[-0.077,-0.039]$
  for respectively $\alpha$ and $\rho$.}
\label{figure:fig11}
\end{figure}

For the seeds-and-plants data the estimates and standard errors
are computed to be
\beqn
1.591\,(0.045),\quad 2.012\, (0.046),\quad -0.836\, (0.132)
\eeqn 
for $\theta_1,\theta_2,\alpha$, with the standard errors in parentheses. 
%
Clearly, the $\alpha$ parameter is nonzero and significantly
negative; this is also borne out of log-likelihood maxima,
where the three-parameter model reaches a value 10.206
higher than the usual Poisson model with independence.
Figure \ref{figure:fig11}  exhibits the confidence curve 
\beq
\cc(\alpha)=\Gamma_1(D(\alpha)),
\label{eq:ccalpha}
\eeq 
with $D(\alpha)=2\{\ell_{\prof}(\hatt\alpha)-\ell_\prof(\alpha)\}$
the deviance function via the log-profile-likelihood
function; see Schweder and Hjort (2016, Chs.~3, 4). 
The 95\% confidence interval for $\alpha$,
obtained from $\{\alpha\colon\cc(\alpha)\le0.95\}$,
is $[-1.046,-0.531]$, somewhat more accurate than
that based directly on $\hatt\alpha\pm1.96\,\se$. 

The observed correlation, for the 958 pairs, is $-0.084$,
whereas the correlation estimated via the fitted
three-parameter model is $-0.061$. It is admittedly
tiny in size, but is significantly negative, 
as demonstrated via analysis of $\alpha$ above. 
Table \ref{table:thedata}, lower part, exhibits
expected values $E_{i,j}=n\hatt f(i,j)$
based on the independent Poisson (left)
and the three-parameter bivariate Poisson models (right).
The Pearson residuals can be computed, for the two models;
these are $P_{i,j}=(N_{i,j}-E_{i,j})/E_{i,j}^{1/2}$,
with $N_{i,j}$ the observed number for cell $(i,j)$.
The fit is drastically better for the 3-parameter correlation
model; the maximum $P_{i,j}$ goes from 3.738 to 2.274,
on their approximately standard normal scale. 
Also, computing the Pearson statistics $K=\sum_{i,j} P_{i,j}^2$
gives 32.601 and 20.422 for the 2- and the 3-parameter model,
for this $6\times6$ table. 

Above we have used $g(x)=\exp(-x)$ for the adjustment
function. We may however also use other values of $t$
in $g(x)=\exp(-tx)$, when constructing and working 
with our model (3.1). The $t$ parameter can be 
viewed as a fine-tuning parameter, perhaps to be 
set in an ad hoc fashion, such as using 
$g(x)=\exp(-x)$ above. It may however also be 
viewed as another statistical parameter, making (\ref{eq:3.1}) 
a four-parameter bivariate Poisson distribution. 
The fuller log-likelihood function becomes 
\beqn
\ell(\theta_1,\theta_2,\alpha,t)
   &=&\sumin \Bigl(\log f(x_i,\theta_1)+\log f(y_i,\theta_2) \\
   & &\qquad 
   +\log[1+\alpha\{g(x_i,t)-a(\theta_1,t)\}
     \{g(y_i,t)-a(\theta_2,t)\}]\Bigr).
\eeqn 
When maximised we find $\hatt t=1.084$, in this 
particular case giving an adjustment function 
$\exp(-1.084\,x)$ close to what we used above,
and the difference is not significant. 
Figure \ref{figure:fig14} (right panel) gives a confidence curve
for the $t$ parameter, via log-likelihood profiling and 
the $\chi^2_1$ transformation of the associated
deviance function, just as for (\ref{eq:ccalpha})
The 90\% interval is rather skewed, from 0.497 to 2.272, 
and the figure also signals a profiled 
log-likelihood function levelling off for increasing~$t$.

In fact, a 96\% interval is $[0.39,\infty)$,
leaving it not implausible that $t$ is very large.
This in turn corresponds to the brutal adjustment
function where $g(x)$ is 1 at zero and 0 for $x\ge 1$. 
This points to the interesting special case of (\ref{eq:3.1}), 
where the bivariate Poisson distribution adjusts
at zero, i.e.~at $(0,0),(1,0),(0,1)$, but then 
all other values in the same fashion: 
\beqn
f(x,y)=f(x,\theta_1)f(y,\theta_2)\times 
\begin{cases}
[1+\alpha \{1-\exp(-\theta_1)\}\{1-\exp(-\theta_2)\}] 
   &{\rm if\ } (x,y)=(0,0), \\
[1-\alpha\{1-\exp(-\theta_1)\}\exp(-\theta_2)] 
   &{\rm if\ } x=0,y\ge1, \\ 
[1-\alpha\exp(-\theta_1)\{1-\exp(-\theta_2)\}]
      &{\rm if\ }  x\ge1,y=0, \\ 
[1+\alpha\exp(-\theta_1)\exp(-\theta_2)] 
      &{\rm if\ }  x\ge1,y\ge1.
\end{cases} 
\eeqn 
This is the particular bivariate Poisson 
matching the right end point of the confidence curve 
$\cc(t)$ displayed in Figure \ref{figure:fig14}, right panel, 
and which here is part of confidence intervals of level 96\% and higher. 



\section{Modelling pairs of correlated binomials}
\label{section:binomialsA}

Suppose $X$ and $Y$ are binomial, say $(n_1,p_1)$ and $(n_2,p_2)$,
but with potential correlation, positive or negative.
If $X$ and $Y$ are counts of a certain event, for two groups,
there might e.g.~be cases where a high $x$ causes disencouragement
for $y$, and vice versa. A model capturing such behaviour,
starting with adjustment functions $g_1(x)$ and $g_2(y)$, is 
\beq
f(x,y)=f_1(x)f_2(y)\{1 + \alpha h_1(x)h_2(y)\},
\quad {\rm with\ }h_1(x)=g_1(x)-a_1,\,h_2(y)=g_2(y)-a_2,
\label{eq:binomial1}
\eeq 
making $h_1(X)$ and $h_2(Y)$ have mean zero under
binomial marginal models $f_1$ and $f_2$.


It is instructive to work through the details for the
adjustment functions $h_1(x)=x/n_1-p_1$ and $h_2(y)=y/n_2-p_2$.
The model is well-defined for a range of $\alpha$ around zero,
say $|\alpha|\le r_0$, but for this particular construction
it turns out that the allowed range of correlation
is small, of the order of $O(1/\sqrt{n_1}+1/\sqrt{n_2})$.
For modelling purposes this would be a too limited
range, unless $n_1$ and $n_2$ are small.

Better choices are the indicator function based
adjustment functions
\beq
h_1(x)=I(x\le x_0)-B(x_0,n_1,p_1)
\quadandquad 
h_2(y)=I(y\le y_0)-B(y_0,n_2,p_2),
\label{eq:binomial2}
\eeq 
for suitable threshold values $x_0,y_0$;
here $B(z,n,p)=\Pr(Z\le z)$ is the binomial c.d.f. We have
\beqn
c_1&=&\max\{ |h_1(x)|\colon x=0,1,\ldots,n_1\}=\max\{B_1(p_1),1-B_1(p_1)\}, \\
c_2&=&\max\{ |h_2(y)|\colon y=0,1,\ldots,n_2\}=\max\{B_2(p_2),1-B_2(p_2)\}, 
\eeqn 
writing $B_1(p_1)=B(x_0,n_1,p_1)$ and similarly for $B_2(p_2)$. 
By (\ref{eq:1.2}), the model (\ref{eq:binomial1}) is bona fide
as long as $|\alpha|<1/(c_1c_2)$. 

\begin{table} 
\begin{small}    
\begin{verbatim}
twenty binomial pairs (x,y) 
    x    y         x    y
   67   65        71   63
   69   61        69   65
   64   71        66   65
   54   71        70   57
   73   66        64   62
   68   64        73   71
   61   66        72   61
   77   66        69   63
   76   62        64   68
   59   66        67   56
\end{verbatim} 
\end{small}
\caption{Twenty simulated binomial pairs $(x_i,y_i)$,
  each from the $\binom(100,p)$, with $p_\true=0.66$, and
  from the model with $\alpha_\true=-1.73$ and
  adjustment functions
  $h_1(x)=I(x\le 66)-B(66,n,p_\true)$ and 
  $h_2(x)=I(y\le 66)-B(66,n,p_\true)$. The pairs
  are shown in Figure \ref{figure:twentypairs}, left panel.}
\label{table:twentypairs} 
\end{table}

\begin{figure}[h]
\centering
\includegraphics[scale=0.36]{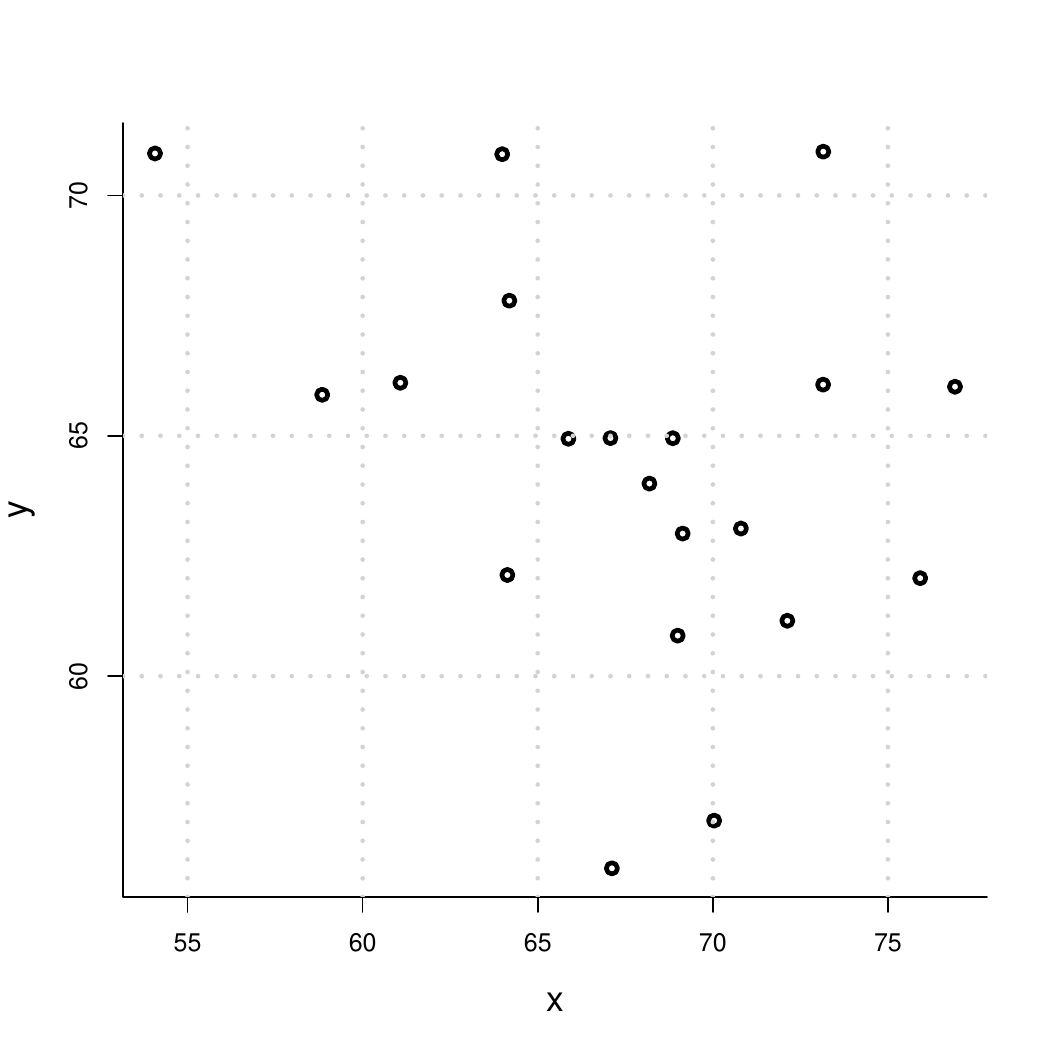}
\includegraphics[scale=0.36]{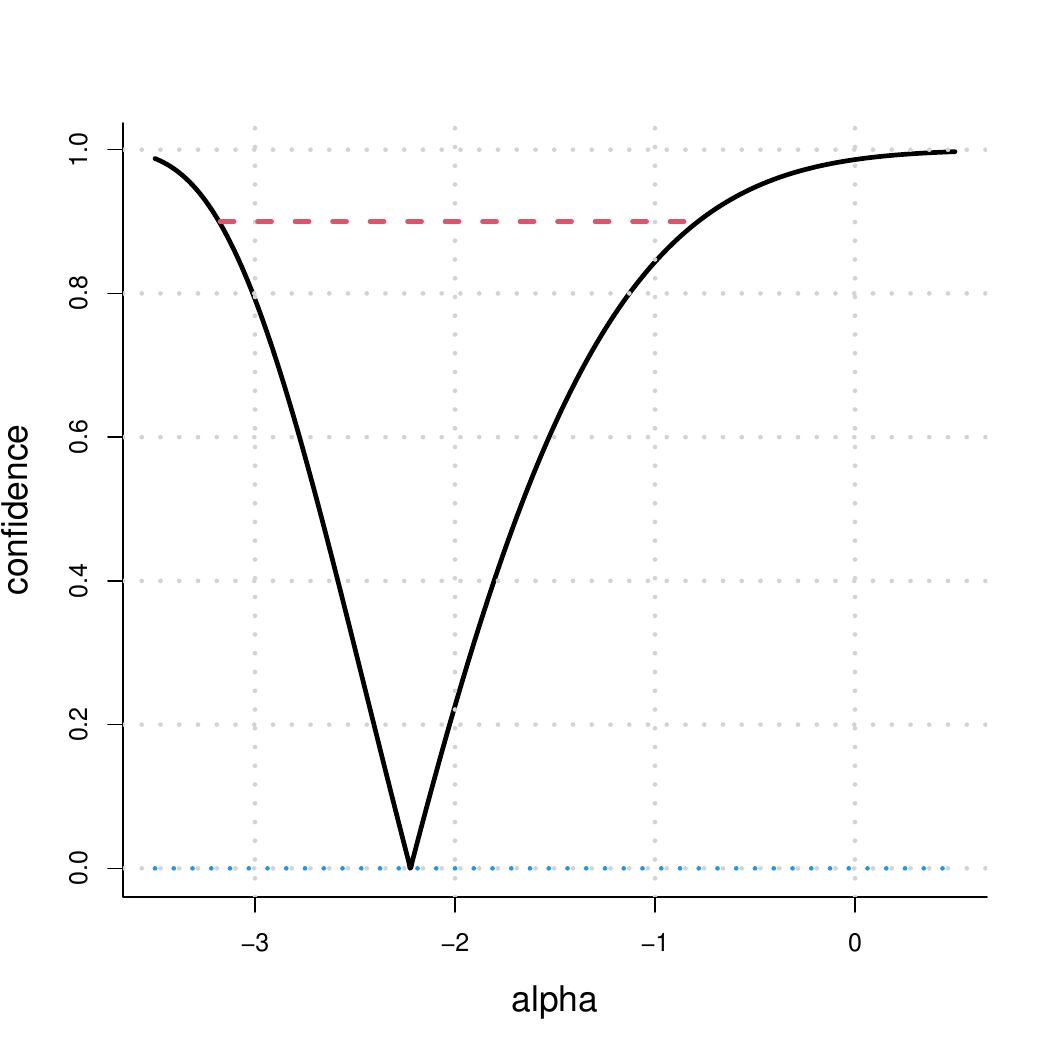}
\caption{Left panel: the $m=20$ binomial pairs $(x_i,y_i)$, 
  with empirical correlation $-0.325$.
  Right panel: confidence curve $\cc(\alpha)$ for the
  adjustment parameter $\alpha$, with point estimate
  $-2.222$; the 90 percent interval is $[-3.137,-0.785]$.}
\label{figure:twentypairs}
\end{figure}

The correlation is $\alpha\nu_1\nu_2/(\sigma_1\sigma_2)$,
by (\ref{eq:1.3}), where we need to work with
\beqn
\nu=\E\,Xh(X)=\E\,X \{I(X\le x_0)-B(p)\}
   =\sum_{x=0}^{x_0} xf(x,p)-np B(p), 
\eeqn 
in terms of a binomial. Writing $X=np+(npq)^{1/2}N_n$,
with variance $npq$ and $N_n$ tending to a standard normal,
we find the following, with two terms cancelling: 
\beqn
\nu&=&\E\, \{np+(npq)^{1/2}N_n\} I(np+(npq)^{1/2}N_n \le x_0)-np B(p) \\
&\doteq& (npq)^{1/2}\,\E\,N_n\,I(N_n\le (x_0-np)/(npq)^{1/2}) \\
&\doteq& -(npq)^{1/2}\phi((x_0-np)/(npq)^{1/2}).  
\eeqn 
As usual, $\phi$ is the standard normal density. Hence
\beqn
\corr(X,Y)\doteq \alpha
   \phi\Bigl({x_0-n_1p_1\over (n_1p_1q_1)^{1/2}}\Bigr)
   \phi\Bigl({y_0-n_2p_2\over (n_2p_2q_2)^{1/2}}\Bigr). 
\eeqn 
If we choose thresholds $x_0,y_0$ close to the means $n_1p_1$
and $n_2p_2$, then the correlation is close to
$\alpha/(2\pi)$, the $c_1,c_2$ above are close to $\half,\half$,
which means that the range for the correlation is about
$|\corr(X,Y)|\le 4/(2\pi)=0.636$. 

To illustrate how the model works, and how the negative
correlation can be assessed via our methods, 
Table \ref{table:twentypairs} gives $m=20$ simulated
binomial pairs $(x_i,y_i)$, each marginal having
the $\binom(100,p_\true)$ structure, with $p_\true=0.66$.
The underlying dependence model is that of (\ref{eq:binomial1}),
with adjustment functions as per (\ref{eq:binomial2}),
with $\alpha_\true=-0.171$; this is $-\half r_0$, where
the allowed range for $\alpha$ is $(-r_0,r_0)$. 
The thresholds $x_0$ and $y_0$
are set to be the rounded value of $\half(\bar x+\bar y)$.
The pairs are plotted in Figure \ref{figure:twentypairs},
left panel. That figure's right panel gives the confidence curve
$\cc(\alpha)$ for the dependence parameter,
computed via the log-likelihood profile function
$\ell_\prof(\alpha)$ in the two-parameter model
with $(p,\alpha)$ the unknowns. As we see, the confidence
intervals of reasonable level capture the underlying
$\alpha_\true$ well. 

\section{Negative binomial correlation across AuditC studies} 
\label{section:binomialsB}

\begin{figure}[h]
\centering
\includegraphics[scale=0.36]{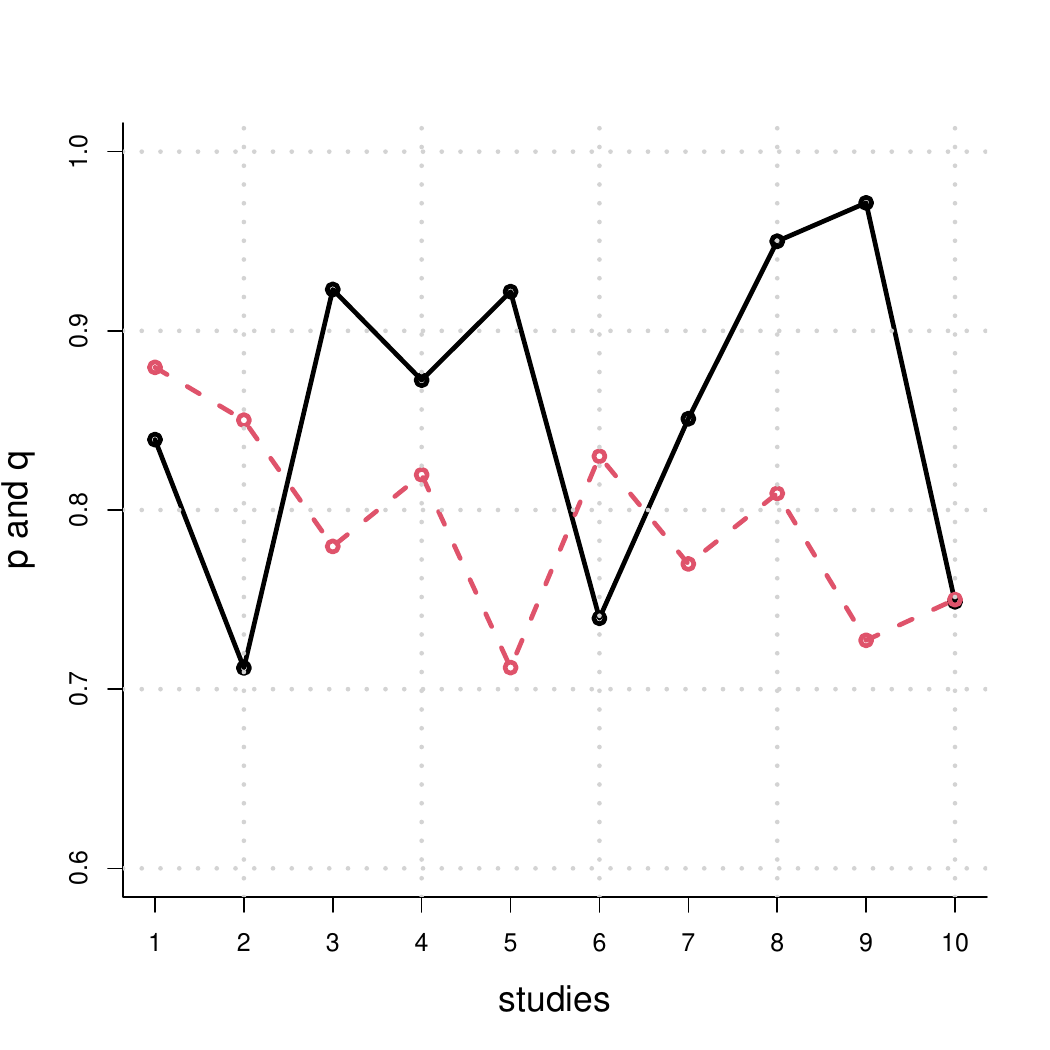}
\includegraphics[scale=0.36]{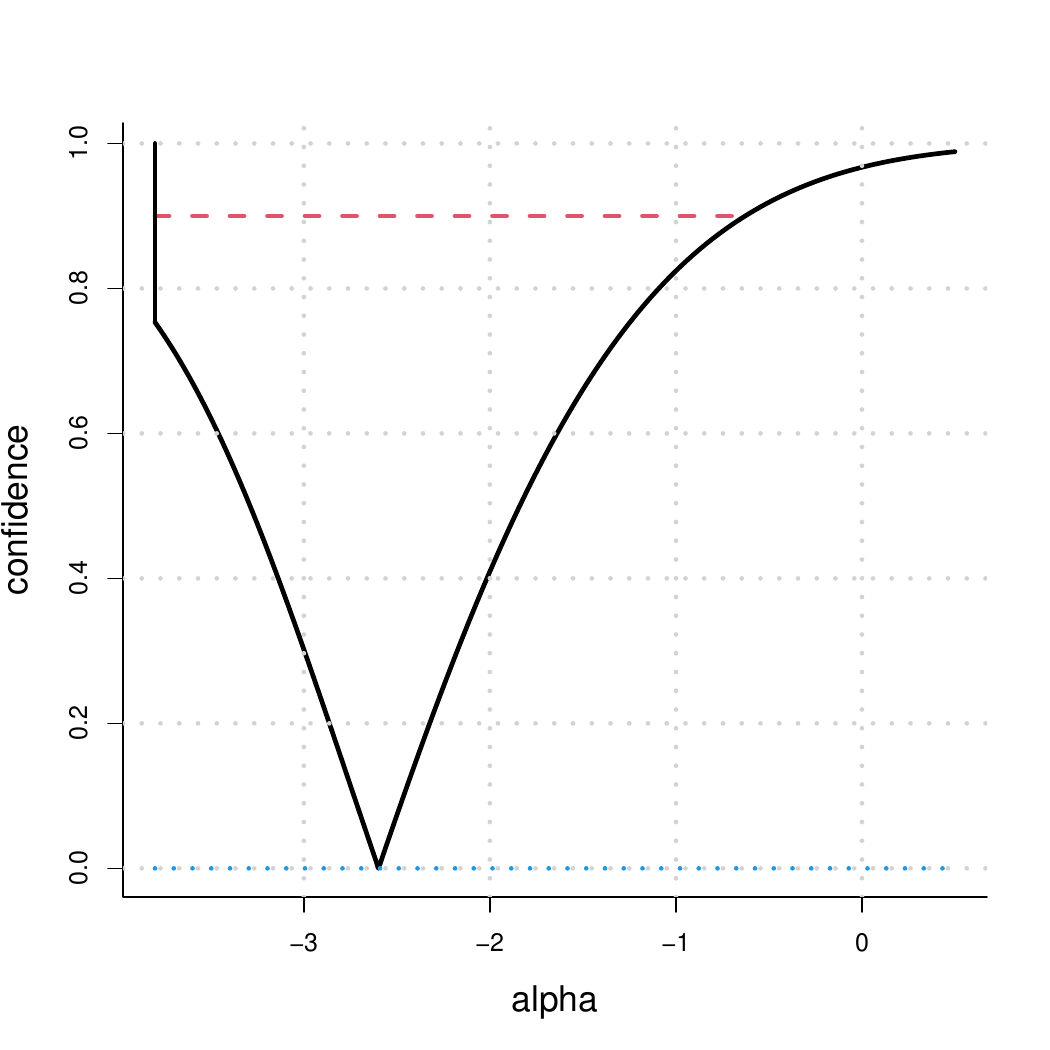}
\caption{For the {\tt mada} Audit-C dataset, with 
  correlated binomial pairs $(x_i,y_i)$ across $m=10$ studies:
  Left panel: the raw estimates $\hatt p_i$ and $\hatt q_i$,
  for respectively true positives in the diseased group
  and true negatives in the non-diseased group.
  Right panel: confidence curve $\cc(\alpha)$
  for the dependence parameter, with point estimate $-2.599$ 
  and 90 percent interval $[-3.80,-0.63]$;
  the allowed range for $\alpha$ does not go to
  the left of $-3.80$.}
\label{figure:mada}
\end{figure}

{\it Audit-C} is a simple three-questions version
of the more general Alcohol Use Disorders Identification Test,
used to identify drinking related risk behaviour
or serious alcoholism.
From the dataset {\tt AuditC} portrayed in the vignette
for the {\tt R} package {\tt mada}, see Doebler (2025),
I have extracted two-times-two table type information
from $m=10$ different studies, pertaining to the quality
of a certain simplified test for the presence of alcohol
problems, compared to a `gold standard' in that field;
see Kriston, H\"olzel, Weiser, Berner, H\"arter (2008) for details. 
The table gives the binomial $X$ of correct `yes',
out of $n_1$ people with alcohol problems,
along with the binomial $Y$ of correct `no',
out of $n_2$ people without such alcohol problems. 
From these binomial counts we also have the
raw estimates $\hatt p_{1,i}=X_i/n_{1,i}$ and $\hatt p_{2,i}=Y_i/n_{2,i}$,
estimates of the underlying probabilities $p$,
for correct diagnosis among the problems people,
and $q$, for correct diagnosis for the not-problem people.
These are shown in Figure \ref{figure:mada}, left panel. 

\begin{table}
\begin{small}
\begin{verbatim}
      X  n1      Y   n2     p      q        
 1   47  56    738  839   0.839  0.880
 2  126 177   1543 1815   0.712  0.850
 3   36  39    276  354   0.923  0.780
 4  130 149    959 1170   0.872  0.820
 5   59  64    136  191   0.922  0.712
 6  142 192   2788 3359   0.740  0.830
 7  137 161    358  465   0.851  0.770
 8   57  60    437  540   0.950  0.809
 9   34  35     56   77   0.971  0.727
10  152 203    264  352   0.749  0.750
\end{verbatim}
\end{small}
\caption{For $m=10$ studies, exhibited are
  $X_i$ from $\binom(n_{1,i},p_i)$
  and $Y_i$ from $\binom(n_{2,i},q_i)$, with $X_i$ the
  number of correct positives among the $n_{1,i}$
  and $Y_i$ the number of correct negatives among the $n_{2,i}$.}
\end{table} 

Among several possible models for these $m$ pairs
of binomials, for the present purposes of assessing
the correlation between them, we use the following
three-parameter model. We use $b(z,n,p)$ and $B(z,n,p)$
for the point probability function and the c.d.f.~of
a binomial $(n,p)$. For study $i$, 
\beqn
f(x_i,y_i)=b(x_i,n_{1,i},p_1)b(y_i,n_{2,i},p_2)
   \{1 + \alpha h_1(x_i)h_2(y_i)\},
\eeqn
with the centred adjustment functions 
\beqn
h_1(x_i)&=&I(x_i\le x_{0,i})-B(x_{0,i},n_{1,i},p_1), \\ 
h_2(y_i)&=&I(y_i\le y_{0,i})-B(y_{0,i},n_{2,i},p_2). 
\eeqn 
The confidence curve $\cc(\alpha)$ is shown in
Figure \ref{figure:mada}, right panel, with point
estimate $\hatt\alpha=-2.60$ and 90 percent interval $[-3.80,-0.63]$
clearly to the left of zero. The point estimate
for the actual correlation is $-0.41$.

\section{Normal marginals, but not binormal}
\label{section:normals}

Let $X$ and $Y$ be standard normal. We shall use (\ref{eq:1.1}) 
to build bivariate models with these standard normals 
as marginals.

\subsection{Normals with adjustment functions.}

Among various similar constructions, consider indeed
\beqn
f(x,y)=\phi(x)\phi(y)[1+\alpha\{\phi(x)-a\}\{\phi(y)-a\}],
\quad {\rm with\ }a=\int\phi(x)^2\,\dd x=\phi(0)/\sqrt{2}. 
\eeqn 
We find that $\max|\phi(x)-a|$ is the same as $a$, so the allowed range
for $\alpha$ is $|\alpha a^2|<1$, i.e.~$|\alpha|<1/a^2=4\pi=12.566$.
The correlation is $\alpha\nu^2$, with
$\nu=\E\,X\{\phi(x)-a\}$, which is zero. So $(X,Y)$ have standard
normal marginals, exhibit dependence, with a wide window
of opportunity for $\alpha$, but with zero correlation. 

In the binormal situation, $\Var\,(Y\midd x)$ is constant in $x$.
Here the dependence structure kicks in differently. From
\beqn
f(y\midd x)=\phi(y) [1 + \alpha\{\phi(x)-a\}\{\phi(y)-a\}] 
\eeqn 
we have
\beqn
\Var\,(Y\midd x)
   =1+\alpha\{\phi(x)-a\}\int y^2\phi(y)\{\phi(y)-a\}\,\dd y
   =1-\half\alpha\, a\{\phi(x)-a\}. 
\eeqn 
For $|x|\le x_0=(\log 2)^{1/2}=0.833$, the $a\{\phi(x)-a\}$
term is positive, and vice versa. 

There are clearly many variants here, with standard normal
marginals, correlation zero, but outside binormality.
A simple construction is
\beqn
f(x,y)=\phi(x)\phi(y)[1 + \alpha \{I(x\le 0)-\half\} \{I(y\le 0)-\half\}],
\eeqn 
where the allowed range for $\alpha$ is seen to be $(-4,4)$.
This model, compared to independence, modifies the probabilities
for the plus-plus and minus-minus quadrants with a certain factor,
and simlarly for the plus-minus and minus-plus with another factor.

Yet another useful construction, for use when fitting
data $(x_i,y_i)$ to a flexible normal marginals model,
say with $(\xi_1,\sigma_1)$ and $(\xi_2,\sigma_2)$
for the two marginals, is to employ adjustment functions
\beqn
h_1(x)&=&\exp\{s_1(x-\xi_1)/\sigma_1-\half t_1(x-\xi_1)^2/\sigma_1^2\}
   -a(s_1,t_1), \\
h_2(y)&=&\exp\{s_2(y-\xi_2)/\sigma_2-\half t_2(y-\xi_2)^2/\sigma_2^2\}
   -a(s_2,t_2), 
\eeqn 
with adjustment function parameters $(s,t)$ either set on prior
grounds or estimated from the data; we take $t_1,t_2$ positive
to ensure boundedness. Here
\beqn
a(s,t)={1\over (1+t)^{1/2}}\exp\Bigl(\half {s^2\over 1+t}\Bigr). 
\eeqn 
Here we do have non-zero correlation. With
\beqn
f(x,y)={1\over \sigma_1}\phi\Bigl({x-\mu_1\over \sigma_1}\Bigr)
       {1\over \sigma_2}\phi\Bigl({y-\mu_2\over \sigma_2}\Bigr)
   \{1+\alpha h_1(x)h_2(y)\}, 
\eeqn 
we have $\corr(X,Y)=\alpha\nu_1\nu_2$, with
\beqn
\nu_1=\int x\phi(x)h_1(x)\,\dd x
   ={\dell a(s_1,t_1)/\dell s_1}
   ={s_1\over (1+t_1)^{3/2}} \exp\Bigl(\half {s_1^2\over 1+t_1}\Bigr), 
\eeqn
and similarly for $\nu_2$.
So this particular bivariate construction
has two parameters for each of the normal marginals,
and then, if we wish for maximal generality,
two plus two plus one for the dependence structure,
i.e.~a quite flexible nine-parameter family.
In applications we might be setting $(s_1,t_1)=(s_2,t_2)$. 
The dependence parameter range is $|\alpha|<1/(c_1c_2)$,
as per (\ref{eq:1.2}), where some analysis reveals that 
\beqn
c_j=\max |h_j(x)|=\max(a_j,g_{j,\max}-a_j),
\quad {\rm with\ } a_j=a(s_j,t_j) {\rm\ and\ }g_{j,\max}=\exp(\half s_j^2/t_j). 
\eeqn 
The correlation is $\alpha\nu_1\nu_2$, via (\ref{eq:1.3}).
For $(s,t)=(1,1)$, for example, the correlation range
is found to be $[-0.25,0.25]$, but it might be bigger for
calibrated choices of $(s,t)$. 

The adjustment function parameters $(s_j,t_j)$ can be estimated
in a two-stage fashion, with the marginal means and variances
estimated first, or via full log-likelihood for the full parameter
vector. See Ko and Hjort (2019) for more on two-stage copula parameter
estimation strategies and how these compare to full maximum likelihood. 

I record here the cleanest bivariate model of this type,
with standard normal marginals and the same adjustment function; then
\beqn
f(x,y)=\phi(x)\phi(y) \bigl[1+\alpha \{\exp(sx-\half tx^2)-a(s,t)\}
   \{\exp(sy-\half ty^2)-a(s,t)\}\bigr], 
\eeqn 
with parameters $s,t,\alpha$ to model the dependence.
There is dependence, though with zero correlation,
is $s=0$, but non-zero correlation, positive of negative,
if $s\not=0$. 

\subsection{Full binormality.}

For the classic binormal, with zero means, unit variances,
and correlation $\rho$, we have via some algebra that 
\beqn
{f_\rho(x,y)\over \phi(x)\phi(y)}
={1\over (1-\rho^2)^{1/2}}
   \exp\Bigl[{\rho\over 1-\rho^2}\{xy-\half\rho(x^2+y^2)\}\Bigr]. 
\eeqn 
For $\rho$ small, this is
\beqn
1 + \rho xy + \rho^2 \{\half x^2y^2-\half(x^2+y^2)+\half\}+O(\rho^3). 
\eeqn
We learn from this that the binormal in essence is close to,
but not equal to, the general recipe (\ref{eq:1.1})
of this article, with adjustment functions $h_1(x)h_2(y)=xy$.
Being unbounded, these are not allowed, however,
though they could be truncated outside say $[-4,4]$,
and then be essentially inside this paper's umbrella. 

\def\poi{{\rm poi}}

\section{Concluding remarks}
\label{section:concluding}

These are a few of my bivariate things.

\smallskip
{\bf A. Exponential marginals.}
The general machinery of (\ref{eq:1.1}) can indeed be
used for modelling dependence between several types
of marginals, though I have focused above on the Poisson
and the binomial cases. A good illustration is to
build dependence around two exponential distributions, say
\beqn
f(x,y)=f(x,\theta_1)f(x,\theta_2)\bigl[1+\alpha\{g(x)-a(\theta_1)\}
  \{g(y)-a(\theta_2)\}\bigr], 
\eeqn 
with $f(x,\theta)=\theta\exp(-\theta x)$. A useful adjustment
function of $g(x)=\exp(-tx)$, for a fine-tuning parameter $t$,
and for which $a(\theta)=\theta/(\theta+t)$. The permissible
$\alpha$ are those with
\beqn
|\alpha|\le {1\over c(\theta_1)c(\theta_2)}
   ={(\theta_1+t)(\theta_2+t)\over \min(\theta_1,t)\min(\theta_2,t)}. 
\eeqn 
There are various constructions in the literature for
positively dependent exponentials, as in
e.g.~Hjort and Khasminskii (1993, Section 6). 
Those authors construct a multivariate model
with exponential marginals, via developing a theory
for the amount of time a diffusion process spends along
different lines. Their probabilistic construction in fact
leads to a full continuous stochastic process 
$V=\{V(a)\colon a>0\}$ where each $V(a)$
is exponential. The correlation function associated 
with that construction is positive, however. 

\smallskip
{\bf B. Lancaster expansions.}
One may contemplate expanding the basic model (\ref{eq:1.1})
to allow second order dependence, say via
\beqn
f(x,y)=f_1(x)f_2(y) \{1+\alpha h_1(x)h_2(y)+\beta h_3(x)h_4(y)\}, 
\eeqn 
for well-chosen new bounded adjustment functions $h_3(x)$ and $h_4(y)$.
With a longer expansion this is reminiscent of what is sometimes
called Lancaster expansions, for classes of bivariate models;
see Lancaster (1969) and Streitberg (1990). 

\smallskip
{\bf C. Trivariate models.}
Lifting the basic (\ref{eq:1.1}) to three or more variables,
as opposed to only caring about pairs $(X,Y)$, is partly straightforward.
A model for triple Poisson counts, for example, would the form
\beqn
f(x,y,z)=f(x,\theta_1)f(y,\theta_2)f(z,\theta_3)
   \bigl[1+\alpha\{g(x)-a(\theta_1)\} \{g(y)-a(\theta_2)\} 
     \{g(z)-a(\theta_3)\}\bigr],
\eeqn 
where $g(x)=\exp(-tx)$ and $a(\theta)=\exp[-\theta\{1-\exp(-t)\}]$. 
In various other setups the step from two to three is a tall one;
see Andreassen (2013) for some such themes. 

\smallskip
{\bf D. Regression for correlated pairs.}
It is clear the the basic (\ref{eq:1.1}) construction
lends itself to regression versions. Suppose
$Y_{i,1}\sim\pois(\mu_{i,1})$ and 
$Y_{i,2}\sim\pois(\mu_{i,2})$, with log-linear mean parameters
$\mu_{i,1}=\exp(x_{i,1}^ \tr\beta_1)$ and 
$\mu_{i,2}=\exp(x_{i,2}^ \tr\beta_2)$ in the usual fashion.
Then a bivariate dependence structure can be appended, via
\beqn
f(y_{i,1},y_{i,2})=\pois(y_{i,1},\mu_{i,1})\pois(y_{i,2},\mu_{i,2})
   \{1+\alpha_i h_{i,1}(y_{i,1})h_{i,2}(y_{i,2})\}, 
\eeqn 
perhaps also with covariates entering the $\alpha_i$ term.
This captures different dependence aspects of such paired Poisson
data compared to e.g.~those in Karlis and Ntzoufras (2005)
and in Hellton et al.~(2020).

\section*{Acknowledgements}

I have appreciated both constructive comments and further
references to related literature from Kristoffer Hellton
and Ingrid Hob\ae k Haff. 

\def\annals{Annals of Statistics}
\def\sjs{Scandinaviaflrcib Journal of Statistics}
\def\jrss{Journal of the Royal Statistical Society}
\def\jasa{Journal of the American Statistical Association}

\section*{References}

\def\reff#1{{\noindent\hangafter=1\hangindent=20pt
  #1\smallskip}}
\parindent20pt
\baselineskip12pt
\parskip4pt

\medskip
\reff{%
Aitchison, J.~and Ho, C. (1989).
The multivariate Poisson-log-normal distribution. 
{\sl Biometrika} {\bf 76}, 643--653.}

\reff{%
Andreassen, C.M. (2013). 
{\sl Models and Inference for Correlated Count Data.}
PhD Dissertation, Department of Mathematics, 
University of Aarhus.}

\reff{%
Claeskens, G.~and Hjort, N.L. (2008).
{\sl Model Selection and Model Averaging.}
Cambridge University Press, Cambridge.}

\reff{%
  Doebler, P. (2025).
  {\tt mada}: Meta-Analysis of Diagnostic Accuracy,
  {\tt R} package version 0.5.12,
  url is {\tt CRAN.R-project.org/package=mada}.}

\reff{%
Edwards, C.B.~and Gurland, J. (1961). 
A class of distributions applicable to accidents.
{\sl\jasa} {\bf 56}, 503--517.}

\reff{%
  Hellton, K.H., Cummings, Vik-Mo, A.U., Nordrehaug, J.E.,
  Aarsland, D., Selbaek, G., and Gill, L.M. (2020).
  The truth behind the zeros: A new approach to principal
  component analysis of the neuropsychiatric inventory.
  {\sl Multivariate Behavioral Research}, {\bf 56}, 70--85.}

\reff{%
Hjort, N.L.~and Khasminskii, R.Z. (1993).
On the time a diffusion process spends a long a line.
{\sl Stochastic Processes and their Applications} {\bf 47},
229--247.}

\reff{%
  Karlis, D.~and Ntzoufras, I. (2005).
  Bivariate Poisson and diagonal inflated bivariate
  Poisson regression models in {\tt R}.
  {\sl Journal of Statistical Software}, {\bf 14}, 1--36.}

\reff{%
  Kriston, L., H{\"o}lzel, L., Weiser. A., Berner. M.,
  and H\"arter, M. (2008).
  Meta-analysis: Are 3 questions enough to detect unhealthy
  alcohol use? {\sl Annals of Internal Medicine}, {\bf 149}, 879--888.}


\reff{%
  Ko, V.~and Hjort, N.L. (2019).
  Copula information criterion for model selection with
  two-stage maximum likelihood estimation.
  Econometrics and Statistics.}

\reff{%
  Ko, V.~and Hjort, N.L. (2019).
  Model robust inference with two-stage maximum likelihood estimation
  for copulas.
  Journal of Multivariate Analysis, 171, 362--381.}

\reff{%
  Ko, V., Hjort, N.L., and Hob\ae k Haff, I. (2019).
  Focused information criteria for copulae.
  Scandinavian Journal of Statistics, 46, 1117--1140.}

\reff{%
  Lancaster, H.O. (1969). {\sl The Chi-Squared Distribution.}
  Wiley, London.} 

\reff{%
Lakshminarayana, J., Pandit, S.N.N, and Rao, K.~Srinivasa (1999).
On a bivariate Poisson distribution. 
{\sl Communications in Statistics -- Theory and methods} {\bf 28},
267--276.} 

\reff{%
  Mikosch, T. (2006).
  Copulas: tales and facts [with discussion and a rejoinder].
  {\sl Extremes}, {\bf 9}, 3--20.}

\reff{%
  Nelson, R. B. (1999).
  {\sl An Introduction to Copulas.}
  Springer-Verlag, Berlin.}

\reff{%
Schweder, T.~and Hjort, N.L. (2016).
{\sl Confidence, Likelihood, Probability.}
Cambridge University Press, Cambridge.}

\reff{%
  Streitberg, B. (1990).
  Lancaster interactions revisited.
  {\sl Annals of Statistics}, {\bf 18}, 1878--1885.}

\reff{%
  Yu, J., Kepner, J.I., and Iyer, R. (2009).
  Exact tests using two correlated binomial variables in
  contemporary cancer clinical trials.
  {\it Biometrical Journal}, {\bf 51}, 899--914.}

\end{document}